\documentclass[11pt]{article}
\usepackage[dvips]{color}
\usepackage{epsfig}
\usepackage{amsmath}
\usepackage{graphicx}
\date{}
\begin{document}
\title{{\bf Bi-directional quantum teleportation of GHZ-like states}}
\author{Leila S. Tabatabaei\,\, and Babak Vakili\thanks{e-mail: b.vakili@iauctb.ac.ir}\\\\{\small {\it Department of Physics,
Central Tehran Branch, Islamic Azad University, Tehran, Iran}}}
\maketitle

\begin{abstract}
In this paper we propose a method through which $n$-qubit states can
simultaneously be bi-directionally transmitted between two users. We
assume that Alice and Bob, the legitimate users, each have a
$n$-qubit GHZ-like state and want to teleport it to the other party.
Also, a four-qubit cluster state plays the role of the quantum
channel of this bi-directional quantum teleportation. The protocol
is based on the method that at first, each user, through a series of
$\mbox{CNOT}$ gates, converts the $n$-qubit state into a single
qubit and some $0$ qubits. Then, by means of the Bell state
measurement and proper operation, the single qubit state is
transferred over the channel between the two sides. By
re-applying the $\mbox{CNOT}$ gates on the transmitted qubits and
auxiliary $0$ states, each user reconstructs the initial GHZ-like
state. Finally, we investige the effects of some kind of noises on the density marix of the channel due to its interaction with the environment
and present a method to protect the channel against the bit-flip error.

\vspace{5mm}\noindent\vspace{0.8mm}\newline Keywords: Quantum teleportation;
Bi-directional teleportation; Noisy channels
\end{abstract}

\section{Introduction}
Quantum teleportation (QT) is probably the most exciting application
of the concept of entanglement in quantum telecommunication theory.
The story is that quantum laws allow a quantum state to be
transferred from one place to another without the state moving in
the physical space between the two places, \cite{Bennet} and
\cite{Bou}. Of course, it should be noted that the desired state
cannot be transferred at a speed faster than the speed of light,
thus causing a violation of the theory of relativity. This is
because in order to complete the QT in the final step, the users
(usually called Alice and Bob) must communicate with each other
through a classical channel. Basically, without such a classical
contact, it is not possible to exchange any information by means of
QT, and as is well known, no classical communication is possible
with a speed faster than the speed of light. For a more complete
view of the various aspects of QT, see \cite{Nils}.

In more recent times, attention was paid to different aspects of the
QT, and in addition to its initial version, where a qubit was
transmitted between two users, more general types were investigated.
As various types of QT, we can mention the following cases
\cite{Types}:

$\bullet$ General QT: As a transmitter, Alice sends quantum states
to Bob through various channels.

$\bullet$ Bi-directional quantum teleportation (BQT): Alice and Bob
are both transmitter and receiver at the same time, sending
different numbers of qubits to each other through different quantum
channels.

$\bullet$ Bi-directional controlled quantum teleportation: In this
type of teleportation , apart from Alice and Bob, there is a third
user, say Charlie, who receives the quantum states from Alice and
transfers them to Bob and vice versa.

$\bullet$ Quantum broadcast: As a transmitter, Alice sends a quantum
state to $n$ users on the other side.

$\bullet$ Quantum multicast: As a transmitter, Alice sends $n$
quantum states to $n$ receivers.

In this paper, we examine a BQT protocol in which the users can
transfer a special $n$-qubit state between themselves through a
four-qubit cluster channel. The method is somehow like to the one
used in \cite{Zhou} to transfer two-qubit state and in \cite{Kazem},
in which an arbitrary number of qubits are bi-directionally
transmitted. However, with a more subtle look at the problem, we
realized that the used protocol works only  for a limited subspace of
the $n$-qubit Hilbert space and not for an arbitrary $n$-qubit state
as claimed in \cite{Kazem}. So, we revised the problem and
considered a $n$-dimensional GHZ-like state for transmission between
two users. The logic of the method is simple: first, by a state-reduction circuit, a series of
$\mbox{CNOT}$ gates on each side, convert the $n$-qubit state into a
single qubit and a number of $|0\rangle$ qubits in the hands of each user.
After these single qubits are exchanged by a four-qubit cluster channel between users,
the original $n$-qubit states are reconstructed by the $\mbox{CNOT}$ gates and some auxiliary $|0\rangle$ qubits in an inverse process.
This is done by a state-reconstruction circuit. Therefore, we conclude that with the presented method,
our $n$-qubit states can be successfully teleported by logical circuits composed of $\mbox{CNOT}$ gates.

In what follows, we first provide a quick review of the teleportation process,
in which the transmitter performs its measurement in the Bell basis, and the receiver,
upon knowing the result of this measurement, applies the appropriate operators on his state at hand to decode the desired state.
We then explained the main problem considered in this article and presented the details of calculations required for
a BQT, in which a four-qubit cluster channel is used to send and receive what we called $n$-qubit GHZ-like state. At the end,
we also took a brief look at the effects of the environment on the channel. We introduce six types
of common noises and in each case, calculate the density matrix of the noisy channel and its fidelity with the original one and provide
a detailed method to fix the bit-flip error.

\section{A brief review of quantum teleportation}
In this section we will have a quick overview of the problem of
QT scheme of a single qubit. Such a process
can be considered either as an one-directional quantum teleportation
of a qubit (from Alice to Bob) or a BQT of a qubit (from Alice to Bob and vice versa).

As proposed in \cite{Bennet}, in QT the main idea is
that Alice wishes to transfer a quantum state to Bob without being
able to deliver the particle in this state directly to him but has a
classical channel linking her to Bob. Suppose that the mentioned
state is a qubit state of the form

\begin{equation}\label{A}
|\Psi\rangle_A=\alpha |0\rangle_A+\beta|1\rangle_A,
\end{equation}where here and also in the rest of the article $|0\rangle$
and $|1\rangle$ are two orthogonal states showing the basis of the
corresponding Hilbert space and complex coefficients $\alpha$ and
$\beta$ satisfying the normalization condition

\begin{equation}\label{B}
|\alpha|^2+|\beta|^2=1.
\end{equation}In order to do the protocol exactly, Alice and Bob
initially share an entangled state, say the Bell state

\begin{equation}\label{C}
|\Phi^{+}\rangle_{AB}=\frac{|00\rangle_{AB}+|11\rangle_{AB}}{\sqrt{2}},
\end{equation}in which Alice has the first qubit and Bob has the
second. So, the process starts with the state $|\Psi\rangle_A\otimes
|\Phi^{+}\rangle_{AB}=|\Psi\rangle_A |\Phi^{+}\rangle_{AB}$. With a
little algebra and keeping the qubits in the same order, this state
may be written as

\begin{equation}\label{D}
|\Psi\rangle_A|\Phi^{+}\rangle_{AB}=\frac{1}{2}\left[|\Phi^{+}\rangle_{AA}|\Psi\rangle_B+|\Psi^{+}\rangle_{AA}\left(X|\Psi\rangle_B\right)+|\Phi^{-}
\rangle_{AA}\left(Z|\Psi\rangle_B\right)+|\Psi^{-}\rangle_{AA}
\left(XZ|\Psi\rangle_B\right)\right],
\end{equation}where

\begin{equation}\label{E}
|\Phi^{\pm}\rangle=\frac{|00\rangle\pm
|11\rangle}{\sqrt{2}},\hspace{5mm}
|\Psi^{\pm}\rangle=\frac{|01\rangle\pm |10\rangle}{\sqrt{2}},
\end{equation}are the Bell states and $X=\left(%
\begin{array}{cc}
  0 & 1 \\
  1 & 0 \\
\end{array}%
\right)$ and $Z=\left(%
\begin{array}{cc}
  1 & 0 \\
  0 & -1 \\
\end{array}%
\right)$ represent the Pauli gates. Now, Alice may perform a Bell
state measurement on the first two qubits. After this measurement
the state of the Alice-Bob system is one of
$|\Phi^{+}\rangle_{AA}|\Psi\rangle_B$,
$|\Psi^{+}\rangle_{AA}\left(X|\Psi\rangle_B\right)$,
$|\Phi^{-}\rangle_{AA}\left(Z|\Psi\rangle_B\right)$ or
$|\Psi^{-}\rangle_{AA} \left(XZ|\Psi\rangle_B\right)$, each with
probability $\frac{1}{4}$. It can be seen that depending on the
measurement result the Bob's state is directly related to the
initial state. So, if Alice informs Bob about her measurement result
through a classical channel, Bob can perform a suitable unitary
transformation on his state at hand in order to obtain the initial
state and this completes the teleportation process, see table 1. It is important to note that
what finally happens is that the particle in Bob's hand becomes the same state as the particle in Alice's hand was in that state from the beginning.

Such a process can be extended to a BQT process in which both Alice and Bob simultaneously
play the role of sender and receiver and send the unknown state to each other. Since in the next section,
we will examine a special type of such teleportation in which a $n$-qubit state is transferred between two parties,
we will not go into the details of such processes here and refer the reader to the relevant references, see for example \cite{Zha} and refs. therein.

\begin{table}\center
\begin{tabular}{|c|c|c|c|}
\hline
Alice measurement results & Bob's states & Bob's operations & Bob's final states \\
\hline
$|\Phi^{+}\rangle$ & $\alpha|0\rangle_B+\beta|1\rangle_B$ & $I$ & $\alpha|0\rangle_B+\beta|1\rangle_B$ \\
$|\Psi^{+}\rangle$ & $\alpha |1\rangle_B+\beta |0\rangle_B$ & $X$ & $\alpha|0\rangle_B+\beta|1\rangle_B$ \\
$|\Phi^{-}\rangle$ & $\alpha |0\rangle_B-\beta |1\rangle_B$ & $Z$ & $\alpha|0\rangle_B+\beta|1\rangle_B$ \\
$|\Psi^{-}\rangle$ & $\alpha|1\rangle_B-\beta|0\rangle_B$ & $ZX$ & $\alpha|0\rangle_B+\beta|1\rangle_B$ \\
\hline
\end{tabular}\caption{The table shows the teleportation protocol of a single qubit from Alice to Bob.}
\end{table}

\section{BQT of $n$-qubit GHZ-like state}
In this section we assume that Alice and Bob want to transfer an
arbitrary number of qubits (indeed the information about the state of the qubits) with pre-determined states
simultaneously. In the beginning of the protocol, suppose the qubit
states belonging to each of these are as follows

\begin{equation}\label{AA}
|\Phi\rangle_{A_1...A_n}=\alpha_0|00...0\rangle+\alpha_1
|11...1\rangle,
\end{equation}and

\begin{equation}\label{AB}
|\Phi\rangle_{B_1...B_n}=\beta_0|00...0\rangle+\beta_1
|11...1\rangle,
\end{equation}
in which $|\alpha_0|^2+|\alpha_1|^2=|\beta_0|^2+|\beta_1|^2=1$. In
the case where $\alpha_0=\alpha_1=1/\sqrt{2}=\beta_0=\beta_1$, these
states are known as the Greenberger-Horne-Zeilinger (GHZ) states
\cite{Green}, which are used to show the entanglement of systems
containing more than two particles. So, let us call the states
mentioned in (\ref{AA}) and (\ref{AB}) the GHZ-like states\footnote{Our definition of what we introduced here as
GHZ-like state may differ from the same definition in other references.}. As the
teleportation channel, we consider the following four-qubit cluster
state shared between Alice and Bob

\begin{equation}\label{AC}
|\Phi\rangle_{a_1a_2b_1b_2}=\frac{1}{2}(|0000\rangle+|0011\rangle+|1100\rangle+|1111\rangle).
\end{equation}Such a channel has already been used in \cite{Zhou} to
BQT of two-qubit states and in \cite{Kazem} for BQT of an arbitrary
number of qubits. Now, let's begin the protocol by writing down the
state of the entire system, that is the product state

\begin{equation}\label{AD}
|\Psi\rangle_{A_1...A_nB_1...B_na_1a_2b_1b_2}=|\Phi\rangle_{A_1...A_n}\otimes
|\Phi\rangle_{B_1...B_n}\otimes |\Phi\rangle_{a_1a_2b_1b_2}.
\end{equation}
At first, we need to find a way to transmit the input qubit
according to the width of the quantum channel. For this purpose,
Alice and Bob's $n$-qubit states must be converted into a tensor
product of entangled $(n-1)$-qubit states and a single $0$-qubit
state. Then, we repeat this for the $(n-1)$-qubit states and this
process continues until the final state becomes a tensor product of
a single superposition state and a $0$-qubit. The mathematical
operations required to perform this process are presented in the
following series of relations

\begin{eqnarray}\label{AE}
\left\{
\begin{array}{ll}
(n-1)-\mbox{CNOT}|\Phi\rangle_{A_1...A_n}\hspace{5mm}
\mapsto\hspace{5mm}|\Phi\rangle_{A_1...A_{n-1}}\otimes
|0\rangle_{A_n},\\\\
(n-2)-\mbox{CNOT}|\Phi\rangle_{A_1...A_{n-1}}\hspace{5mm}
\mapsto\hspace{5mm}|\Phi\rangle_{A_1...A_{n-2}}\otimes
|0\rangle_{A_{n-1}},\\\\
...\\\\
\mbox{Toffoli}|\Phi\rangle_{A_1A_2A_3}\hspace{5mm}\mapsto
\hspace{5mm}|\Phi\rangle_{A_1A_2}\otimes|0\rangle_{A_3},\\\\
\mbox{CNOT}|\Phi\rangle_{A_1A_2}\hspace{5mm}\mapsto
\hspace{5mm}|\Phi\rangle_{A_1}\otimes |0\rangle_{A_2}.
\end{array}
\right.
\end{eqnarray}
By performing similar operations on Bob's $n$-qubit state
$|\Phi\rangle_{B_1...B_n}$, it also becomes
$|\Phi\rangle_{B_1}\otimes |0\rangle_{B_2}$. Now, our problem is
reduced to the transmission of the states

\begin{eqnarray}\label{AF}
\left\{
\begin{array}{ll}
|\Phi\rangle_{A_1}=\left(\alpha_0 |0\rangle+\alpha_1 |1\rangle\right)_{A_1},\\\\
|\Phi\rangle_{B_1}=\left(\beta_0 |0\rangle+\beta_1
|1\rangle\right)_{B_1},
\end{array}
\right.
\end{eqnarray}through the channel mentioned in (\ref{AC}). We may
now rewrite the state of the total system as

\begin{eqnarray}\label{AFF}
|\Psi\rangle_{A_1B_1a_1a_2b_1b_2}=|\Phi\rangle_{A_1}\otimes
|\Phi\rangle_{B_1}\otimes |\Phi\rangle_{a_1a_2b_1b_2}&=&\nonumber\\
\left(\alpha_0 |0\rangle+\alpha_1 |1\rangle\right)_{A_1}\otimes
\left(\beta_0 |0\rangle+\beta_1 |1\rangle\right)_{B_1}\otimes
\frac{1}{2}\left(|0000\rangle+|0011\rangle+|1100\rangle+|1111\rangle
\right)_{a_1a_2b_1b_2}=\nonumber\\
\alpha_0\beta_0|00\rangle_{A_1a_1}|00\rangle_{a_2b_2}|00\rangle_{B_1b_1}+\alpha_0\beta_0|00\rangle_{A_1a_1}|01\rangle_{a_2b_2}|01\rangle_{B_1b_1}+\alpha_0
\beta_0|01\rangle_{A_1a_1}|10\rangle_{a_2b_2}|00\rangle_{B_1b_1}+
\nonumber\\
\alpha_0\beta_0|01\rangle_{A_1a_1}|11\rangle_{a_2b_2}|01\rangle_{B_1b_1}+\alpha_0\beta_1|00\rangle_{A_1a_1}|00\rangle_{a_2b_2}|10\rangle_{B_1b_1}+\alpha_0\beta_1
|00\rangle_{A_1a_1}|01\rangle_{a_2b_2}|11\rangle_{B_1b_1}+\nonumber
\\ \alpha_0\beta_1|01\rangle_{A_1a_1}|10\rangle_{a_2b_2}|10\rangle_{B_1b_1}+\alpha_0\beta_1|01\rangle_{A_1a_1}|11\rangle_{a_2b_2}|11\rangle_{B_1b_1}+
\alpha_1\beta_0|10\rangle_{A_1a_1}|00\rangle_{a_2b_2}|00\rangle_{B_1b_1}+\nonumber
\\ \alpha_1\beta_0
|10\rangle_{A_1a_1}|01\rangle_{a_2b_2}|01\rangle_{B_1b_1}+\alpha_1\beta_0|11\rangle_{A_1a_1}|10\rangle_{a_2b_2}|00\rangle_{B_1b_1}+
\alpha_1\beta_0|11\rangle_{A_1a_1}|11\rangle_{a_2b_2}|01\rangle_{B_1b_1}+\nonumber
\\
\alpha_1\beta_1|10\rangle_{A_1a_1}|00\rangle_{a_2b_2}|10\rangle_{B_1b_1}+\alpha_1\beta_1|10\rangle_{A_1a_1}|01\rangle_{a_2b_2}|11\rangle_{B_1b_1}+
\alpha_1\beta_1|11\rangle_{A_1a_1}|10\rangle_{a_2b_2}|10\rangle_{B_1b_1}+\nonumber
\\
\alpha_1\beta_1|11\rangle_{A_1a_1}|11\rangle_{a_2b_2}|11\rangle_{B_1b_1}.
\end{eqnarray}
In terms of the Bell basis (\ref{E}), the above
relation can be written as

\begin{eqnarray}\label{AG}
|\Psi\rangle_{A_1B_1a_1a_2b_1b_2}&=&\nonumber\\
\frac{1}{2} \left[\left(|\Phi^+\rangle+|\Phi^-\rangle\right)\left(|\Phi^+\rangle+|\Phi^-\rangle\right)\right]\alpha_0\beta_0|00\rangle
+\frac{1}{2} \left[\left(|\Phi^+\rangle+|\Phi^-\rangle\right)\left(|\Phi^+\rangle-|\Phi^-\rangle\right)\right]\alpha_0\beta_1|01\rangle+\nonumber\\
\frac{1}{2}
\left[\left(|\Phi^+\rangle-|\Phi^-\rangle\right)\left(|\Phi^+\rangle+|\Phi^-\rangle\right)\right]\alpha_1\beta_0|10\rangle
+\frac{1}{2}
\left[\left(|\Phi^+\rangle-|\Phi^-\rangle\right)\left(|\Phi^+\rangle-|\Phi^-\rangle\right)\right]\alpha_1\beta_1|11\rangle+\nonumber\\
\frac{1}{2}
\left[\left(|\Phi^+\rangle+|\Phi^-\rangle\right)\left(|\Psi^+\rangle+|\Psi^-\rangle\right)\right]\alpha_0\beta_0|01\rangle
+\frac{1}{2}
\left[\left(|\Phi^+\rangle+|\Phi^-\rangle\right)\left(|\Psi^+\rangle-|\Psi^-\rangle\right)\right]\alpha_0\beta_1|00\rangle+\nonumber\\
\frac{1}{2}
\left[\left(|\Phi^+\rangle-|\Phi^-\rangle\right)\left(|\Psi^+\rangle+|\Phi^-\rangle\right)\right]\alpha_1\beta_0|11\rangle
+\frac{1}{2}
\left[\left(|\Phi^+\rangle-|\Phi^-\rangle\right)\left(|\Psi^+\rangle-|\Psi^-\rangle\right)\right]\alpha_1\beta_1|10\rangle+\nonumber\\
\frac{1}{2}
\left[\left(|\Psi^+\rangle+|\Psi^-\rangle\right)\left(|\Phi^+\rangle+|\Phi^-\rangle\right)\right]\alpha_0\beta_0|10\rangle
+\frac{1}{2}
\left[\left(|\Psi^+\rangle+|\Psi^-\rangle\right)\left(|\Phi^+\rangle-|\Phi^-\rangle\right)\right]\alpha_0\beta_1|11\rangle+\nonumber\\
\frac{1}{2}
\left[\left(|\Psi^+\rangle-|\Psi^-\rangle\right)\left(|\Phi^+\rangle-|\Phi^-\rangle\right)\right]\alpha_1\beta_0|00\rangle
+\frac{1}{2}
\left[\left(|\Psi^+\rangle-|\Psi^-\rangle\right)\left(|\Phi^+\rangle-|\Phi^-\rangle\right)\right]\alpha_1\beta_1|01\rangle+\nonumber\\
\frac{1}{2}
\left[\left(|\Psi^+\rangle+|\Psi^-\rangle\right)\left(|\Psi^+\rangle+|\Psi^-\rangle\right)\right]\alpha_0\beta_0|11\rangle
+\frac{1}{2}
\left[\left(|\Psi^+\rangle+|\Psi^-\rangle\right)\left(|\Psi^+\rangle-|\Psi^-\rangle\right)\right]\alpha_0\beta_1|10\rangle+\nonumber\\
\frac{1}{2}
\left[\left(|\Psi^+\rangle-|\Psi^-\rangle\right)\left(|\Psi^+\rangle+|\Psi^-\rangle\right)\right]\alpha_1\beta_0|01\rangle
+\frac{1}{2}
\left[\left(|\Psi^+\rangle-|\Psi^-\rangle\right)\left(|\Psi^+\rangle-|\Psi^-\rangle\right)\right]\alpha_1\beta_1|00\rangle,
\end{eqnarray}
where after some rearranging takes the form

\begin{eqnarray}\label{AH}
|\Psi\rangle_{A_1B_1a_1a_2b_1b_2}=|\Phi^+\rangle_{A_1a_1}|\Phi^+\rangle_{B_1b_1}|\eta^1\rangle_{a_2b_2}+|\Phi^+
\rangle_{A_1a_1}|\Phi^-\rangle_{B_1b_1}|\eta^2\rangle_{a_2b_2}+\nonumber
\\|\Phi^-\rangle_{A_1a_1}|\Phi^+\rangle_{B_1b_1}|\eta^3\rangle_{a_2b_2}+|\Phi^-\rangle_{A_1a_1}|\Phi^-\rangle_{B_1b_1}|\eta^4\rangle_{a_2b_2}+\nonumber
\\|\Phi^+\rangle_{A_1a_1}|\Psi^+\rangle_{B_1b_1}|\eta^5\rangle_{a_2b_2}+|\Phi^+\rangle_{A_1a_1}|\Psi^-\rangle_{B_1b_1}|\eta^6\rangle_{a_2b_2}+\nonumber
\\|\Phi^-\rangle_{A_1a_1}|\Psi^+\rangle_{B_1b_1}|\eta^7\rangle_{a_2b_2}+|\Phi^-\rangle_{A_1a_1}|\Psi^-\rangle_{B_1b_1}|\eta^8\rangle_{a_2b_2}+\nonumber
\\|\Psi^+\rangle_{A_1a_1}|\Phi^+\rangle_{B_1b_1}|\eta^9\rangle_{a_2b_2}+|\Psi^+\rangle_{A_1a_1}|\Phi^-\rangle_{B_1b_1}|\eta^{10}\rangle_{a_2b_2}+\nonumber
\\|\Psi^-\rangle_{A_1a_1}|\Phi^+\rangle_{B_1b_1}|\eta^{11}\rangle_{a_2b_2}+|\Psi^-\rangle_{A_1a_1}|\Phi^-\rangle_{B_1b_1}|\eta^{12}\rangle_{a_2b_2}+\nonumber
\\|\Psi^+\rangle_{A_1a_1}|\Psi^+\rangle_{B_1b_1}|\eta^{13}\rangle_{a_2b_2}+|\Psi^+\rangle_{A_1a_1}|\Psi^-\rangle_{B_1b_1}|\eta^{14}\rangle_{a_2b_2}+\nonumber
\\|\Psi^-\rangle_{A_1a_1}|\Psi^+\rangle_{B_1b_1}|\eta^{15}\rangle_{a_2b_2}+|\Psi^-\rangle_{A_1a_1}|\Psi^-\rangle_{B_1b_1}|\eta^{16}\rangle_{a_2b_2},
\end{eqnarray}
in which the states $|\eta^i\rangle_{a_2b_2}$ ($i=1...16$) are
defined as

\begin{eqnarray}\label{AI}
\left\{
\begin{array}{ll}
|\eta^1\rangle_{a_2b_2}=\alpha_0\beta_0|00\rangle+\alpha_0\beta_1|01\rangle+\alpha_1\beta_0|10\rangle+\alpha_1\beta_1|11\rangle,\\
|\eta^2\rangle_{a_2b_2}=\alpha_0\beta_0|00\rangle-\alpha_0\beta_1|01\rangle+\alpha_1\beta_0|10\rangle-\alpha_1\beta_1|11\rangle,\\

|\eta^3\rangle_{a_2b_2}=\alpha_0\beta_0|00\rangle+\alpha_0\beta_1|01\rangle-\alpha_1\beta_0|10\rangle-\alpha_1\beta_1|11\rangle,\\

|\eta^4\rangle_{a_2b_2}=\alpha_0\beta_0|00\rangle-\alpha_0\beta_1|01\rangle-\alpha_1\beta_0|10\rangle+\alpha_1\beta_1|11\rangle,\\\\

|\eta^5\rangle_{a_2b_2}=\alpha_0\beta_0|01\rangle+\alpha_0\beta_1|00\rangle+\alpha_1\beta_0|11\rangle+\alpha_1\beta_1|10\rangle,\\

|\eta^6\rangle_{a_2b_2}=\alpha_0\beta_0|01\rangle-\alpha_0\beta_1|00\rangle+\alpha_1\beta_0|11\rangle-\alpha_1\beta_1|10\rangle,\\

|\eta^7\rangle_{a_2b_2}=\alpha_0\beta_0|01\rangle+\alpha_0\beta_1|00\rangle-\alpha_1\beta_0|11\rangle-\alpha_1\beta_1|10\rangle,\\

|\eta^8\rangle_{a_2b_2}=\alpha_0\beta_0|01\rangle-\alpha_0\beta_1|00\rangle-\alpha_1\beta_0|11\rangle+\alpha_1\beta_1|10\rangle,\\\\

|\eta^9\rangle_{a_2b_2}=\alpha_0\beta_0|10\rangle+\alpha_0\beta_1|11\rangle+\alpha_1\beta_0|00\rangle+\alpha_1\beta_1|01\rangle,\\

|\eta^{10}\rangle_{a_2b_2}=\alpha_0\beta_0|10\rangle-\alpha_0\beta_1|11\rangle+\alpha_1\beta_0|00\rangle-\alpha_1\beta_1|01\rangle,\\

|\eta^{11}\rangle_{a_2b_2}=\alpha_0\beta_0|10\rangle+\alpha_0\beta_1|11\rangle-\alpha_1\beta_0|00\rangle-\alpha_1\beta_1|01\rangle,\\

|\eta^{12}\rangle_{a_2b_2}=\alpha_0\beta_0|10\rangle-\alpha_0\beta_1|11\rangle-\alpha_1\beta_0|00\rangle+\alpha_1\beta_1|01\rangle,\\\\

|\eta^{13}\rangle_{a_2b_2}=\alpha_0\beta_0|11\rangle+\alpha_0\beta_1|10\rangle+\alpha_1\beta_0|01\rangle+\alpha_1\beta_1|00\rangle,\\

|\eta^{14}\rangle_{a_2b_2}=\alpha_0\beta_0|11\rangle-\alpha_0\beta_1|10\rangle+\alpha_1\beta_0|01\rangle-\alpha_1\beta_1|00\rangle,\\

|\eta^{15}\rangle_{a_2b_2}=\alpha_0\beta_0|11\rangle+\alpha_0\beta_1|10\rangle-\alpha_1\beta_0|01\rangle-\alpha_1\beta_1|00\rangle,\\

|\eta^{16}\rangle_{a_2b_2}=\alpha_0\beta_0|11\rangle-\alpha_0\beta_1|10\rangle-\alpha_1\beta_0|01\rangle+\alpha_1\beta_1|00\rangle.\\

\end{array}
\right.
\end{eqnarray}Now, in the measurement step, Alice puts her input information, (that is $A_1$), on the
channel $a_1$, and Bob do the same with $B_1$ and puts it on the
channel $b_1$. Then, both of them perform their measurements in the
Bell basis. As a result of these measurements, the remaining states,
i.e. the states of the channel $a_2b_2$, are collapsed into sixteen
modes (which we have denoted them by $|\eta^i\rangle$) which are
responsible for transmitting the information to the users. Now, it is
time for Alice and Bob to inform each other about the results of
their measurements through a classical channel. Notice that their
initial qubit states are completely destroyed upon their
measurements. This makes quantum teleportation consistent with the
no-cloning theorem. After each of them receives the signal sent from
the other side, he (she) can reconstruct her (his) initial state by
applying the decoding process, that is, applying an appropriate
operator on the collapsed state. The operations required for Alice
and Bob to restore the initial states after knowing the results of the
measurements are shown in the table 2. Again, it should be noted that according to the
third column of this table, the particles in the hands of Alice and Bob have become the state that the other party intended to send.
This is why the states like $\alpha_0|0\rangle+\alpha_1|1\rangle$ and $\beta_0|0\rangle+\beta_1|1\rangle$ have indices $b_2$ and $a_2$ respectively.
This important point is not observed in \cite{Kazem}.

\begin{table}\center
\begin{tabular}{|c|c|c|c|c|}
\hline
Alice's results& Bob's results & collapsed states ($|\eta^i\rangle_{a_2b_2})$ & Alice's operations & Bob's operations
\\
\hline $|\Phi^+\rangle_{A_1a_1}$ & $|\Phi^+\rangle_{B_1b_1}$ &
$\left(\alpha_0|0\rangle+\alpha_1|1\rangle\right)_{b_2}\otimes \left(\beta_0|0\rangle+\beta_1|1\rangle\right)_{a_2}$& I & I
\\ \\
$|\Phi^+\rangle_{A_1a_1}$ & $|\Phi^-\rangle_{B_1b_1}$ & $\left(\alpha_0|0\rangle+\alpha_1|1\rangle\right)_{b_2}
\otimes \left(\beta_0|0\rangle-\beta_1|1\rangle\right)_{a_2}$ & Z & I
\\ \\
$|\Phi^+\rangle_{A_1a_1}$ & $|\Psi^+\rangle_{B_1b_1}$ &
$\left(\alpha_0|0\rangle-\alpha_1|1\rangle\right)_{b_2} \otimes
\left(\beta_0|0\rangle+\beta_1|1\rangle\right)_{a_2}$ & I & Z \\ \\
$|\Phi^+\rangle_{A_1a_1}$ & $|\Psi^-\rangle_{B_1b_1}$ &
$\left(\alpha_0|0\rangle-\alpha_1|1\rangle\right)_{b_2} \otimes
\left(\beta_0|0\rangle-\beta_1|1\rangle\right)_{a_2}$ & Z & Z \\ \\
$|\Phi^-\rangle_{A_1a_1}$ & $|\Phi^+\rangle_{B_1b_1}$ & $\left(\alpha_0|0\rangle+\alpha_1|1\rangle\right)_{b_2}
\otimes \left(\beta_0|1\rangle+\beta_1|0\rangle\right)_{a_2}$ & X & I
\\\\
$|\Phi^-\rangle_{A_1a_1}$ & $|\Phi^-\rangle_{B_1b_1}$ & $\left(\alpha_0|0\rangle+\alpha_1|1\rangle\right)_{b_2}
\otimes \left(\beta_0|1\rangle-\beta_1|0\rangle\right)_{a_2}$ & XZ & I
\\\\
$|\Phi^-\rangle_{A_1a_1}$ & $|\Psi^+\rangle_{B_1b_1}$ &
$\left(\alpha_0|0\rangle-\alpha_1|1\rangle\right)_{b_2}
\otimes \left(\beta_0|1\rangle+\beta_1|0\rangle\right)_{a_2}$ & X & Z
\\\\
$|\Phi^-\rangle_{A_1a_1}$ & $|\Psi^-\rangle_{B_1b_1}$ &
$\left(\alpha_0|0\rangle-\alpha_1|1\rangle\right)_{b_2} \otimes
\left(\beta_0|1\rangle-\beta_1|0\rangle\right)_{a_2}$ & XZ & Z
\\\\
$|\Psi^+\rangle_{A_1a_1}$ & $|\Phi^+\rangle_{B_1b_1}$ &
$\left(\alpha_0|1\rangle+\alpha_1|0\rangle\right)_{b_2}
\otimes \left(\beta_0|0\rangle+\beta_1|1\rangle\right)_{a_2}$ & I & X
\\\\
$|\Psi^+\rangle_{A_1a_1}$ & $|\Phi^-\rangle_{B_1b_1}$ &
$\left(\alpha_0|1\rangle+\alpha_1|0\rangle\right)_{b_2}
\otimes \left(\beta_0|0\rangle-\beta_1|1\rangle\right)_{a_2}$ & Z & X
\\\\
$|\Psi^+\rangle_{A_1a_1}$ & $|\Psi^+\rangle_{B_1b_1}$ &
$\left(\alpha_0|1\rangle-\alpha_1|0\rangle\right)_{b_2}
\otimes \left(\beta_0|0\rangle+\beta_1|1\rangle\right)_{a_2}$ & I & XZ
\\\\
$|\Psi^+\rangle_{A_1a_1}$ & $|\Psi^-\rangle_{B_1b_1}$ &
$\left(\alpha_0|1\rangle-\alpha_1|0\rangle\right)_{b_2}
\otimes \left(\beta_0|0\rangle-\beta_1|1\rangle\right)_{a_2}$ & Z & XZ
\\\\
$|\Psi^-\rangle_{A_1a_1}$ & $|\Phi^+\rangle_{B_1b_1}$ &
$\left(\alpha_0|1\rangle+\alpha_1|0\rangle\right)_{b_2} \otimes
\left(\beta_0|1\rangle+\beta_1|0\rangle\right)_{a_2}$ & X & X
\\\\
$|\Psi^-\rangle_{A_1a_1}$ & $|\Phi^-\rangle_{B_1b_1}$ &
$\left(\alpha_0|1\rangle+\alpha_1|0\rangle\right)_{b_2}
\otimes \left(\beta_0|1\rangle-\beta_1|0\rangle\right)_{a_2}$ & XZ & X
\\\\
$|\Psi^-\rangle_{A_1a_1}$ & $|\Psi^+\rangle_{B_1b_1}$ &
$\left(\alpha_0|1\rangle-\alpha_1|0\rangle\right)_{b_2}
\otimes \left(\beta_0|1\rangle+\beta_1|0\rangle\right)_{a_2}$ & X & XZ
\\\\
$|\Psi^-\rangle_{A_1a_1}$ & $|\Psi^-\rangle_{B_1b_1}$ &
$\left(\alpha_0|1\rangle-\alpha_1|0\rangle\right)_{b_2}
\otimes \left(\beta_0|1\rangle-\beta_1|0\rangle\right)_{a_2}$ & XZ &  XZ \\
\hline
\end{tabular}\caption{The table shows the BQT protocol of arbitrary number of qubits between Alice and Bob.}
\end{table}

At the end of this step, Alice and Bob each have a superposition of
the states $|0\rangle$ and $|1\rangle$. They are
$\alpha_0|0\rangle+\alpha_1|1\rangle$ and
$\beta_0|0\rangle+\beta_1|1\rangle$, which are on Bob's and Alice's
sides, respectively. What remains to complete the BQT process we are
considering, is to transform these superposition states into the
initial states (\ref{AA}) and (\ref{AB}). This can be done with the
help of $\mbox{CNOT}$ gates and auxiliary qubits $|0\rangle$.
In fact, in a way, the series of the relations (\ref{AE}) must be
passed in the reverse direction. So, for example, Bob can perform
the following operations on the state he owns

\begin{eqnarray}\label{AJ}
\left\{
\begin{array}{ll}
\mbox{CNOT}\left(\alpha_0|0\rangle+\alpha_1|1\rangle\right)\otimes
|0\rangle\hspace{5mm}
\mapsto\hspace{5mm}\alpha_0|00\rangle+\alpha_1|11\rangle,\\\\
\mbox{Toffoli}\left(\alpha_0|00\rangle+\alpha_1|11\rangle\right)\otimes|0\rangle\hspace{5mm}
\mapsto\hspace{5mm}\alpha_0|000\rangle+\alpha_1|111\rangle,\\\\
...\\\\
(n-2)-\mbox{CNOT}\left(\alpha_0|\underbrace{00...0}_{n-2}\rangle+\alpha_1|\underbrace{11...1}_{n-2}\rangle\right)\otimes
|0\rangle\hspace{5mm}\mapsto \hspace{5mm}\alpha_0|\underbrace{00...0}_{n-1}\rangle+\alpha_1|\underbrace{11...1}_{n-1}\rangle,\\\\
(n-1)-\mbox{CNOT}\left(\alpha_0|\underbrace{00...0}_{n-1}\rangle+\alpha_1|\underbrace{11...1}_{n-1}\rangle\right)\otimes
|0\rangle\hspace{5mm}\mapsto
\hspace{5mm}\alpha_0|\underbrace{00...0}_{n}\rangle+\alpha_1|\underbrace{11...1}_{n}\rangle,
\end{array}
\right.
\end{eqnarray}in which all the states belong to Bob. A similar operation can be performed by Alice
to get $\beta_0|00...0\rangle+\beta_1|11...1\rangle$, which finally
completes the process.

Let's conclude this section by examining whether
the above protocol also work for teleportation of other kinds of $n$-qubit states. For the states of three qubits systems
there are two inequivalent classes of entangled states: GHZ-class $\frac{1}{\sqrt{2}}\left(|000\rangle+|111\rangle\right)$,
and W-class $\frac{1}{\sqrt{3}}\left(|100\rangle+|010\rangle+|001\rangle\right)$, \cite{Dur}, which con not be converted to each other
by stochastic local operations and classical communication \cite {CH}. Their generalization to $n$-qunit states for GHZ are what we already
have considered in (\ref{AA}) and (\ref{AB}) for which we designed a BQT mechanism. Also, a generalized $n$-qubit W-like state can be written as
\begin{equation}\label{AJ0}
|gW\rangle_n=a_1|1000...0\rangle+a_2|0100...0\rangle+a_3|0010...0\rangle+...+a_n|0000...1\rangle,
\end{equation}in which the number of terms in the superposition is equal to the number of qubits and $\sum_{i=1}^n |a_i|^2=1$. Choosing
of such states as the channels in various types of teleportation processes has been widely studied in literature. Indeed unlike GHZ states,
generalized W-states are not always suitable for teleportation's channel. However, there are some restrictive conditions under which these states
may be used as quantum resources to teleport unknown single and
a limited class of multiqubit states, see \cite{Pati} and the references therein. Now, let's go back to the main question,
whether a W-like state itself can be bi-directionally teleported with a protocol similar to what we described in this section.
In general, to teleport an arbitrary $n$-qubit state bi-directionally, one needs a $4n$-qubit channel. However, if the number of qubits required for teleport
can be reduced by mechanisms similar to the relations (\ref{AE}), the number of qubits in the transmission channel will also be reduced. A generalized $2$-qubit
W-like state has the form $|gW\rangle_2=a_1|10\rangle+a_2|01\rangle$, which can be converted to a GHZ-like state by applying an
$X$ gate on the first qubit. Thus, for this simple case our protocol works to teleport W-like state bi-directionally.
Apart from this trivial case, it does not seem possible to condense the generalized W-like states into single qubit states, but rather may be mapped into a
$k$-qubit system where $k=[\log_2 n]+1$, ($[...]$ denotes the integer part). If such condensation is possible, the required channel is different from
one we have introduced in our model. Further investigation of this issue is beyond the scope of this article.

\section{Notes on noisy channel}
As we know, one of the important issues related to the teleportation
is the interaction of the channel with its environment, which leads to the noise in the channel.
Noisy channels disrupt the teleportation process and cause errors in it. One of the most common methods
of investigating the interaction of the system with the environment is the Kraus formalism \cite{Nils} and \cite {Mc},
based on which the density matrix of the principal system will be mapped as

\begin{equation}\label{AL}
\rho \mapsto \varrho=\sum_k E_k \rho E_k^\dag,
\end{equation}where $E_k$ are known as Kraus operators. This map (or to be more precise, supermap)
is a special case of more general maps called quantum operations, that transform linear operators
(not necessarily density operators) of a vector space to linear operators of another one.
The condition for ${\cal F}$  to be a quantum operation is that it satisfies some axioms, one of which is that ${\cal F}$ never increases the trace \cite {Mc}.
In Kraus representation, the trace-decreasing condition imposes the relation $\sum_k E_kE_k^\dag\leq I$. However, most quantum operations preserve the trace
\footnote {An important exception is the process associated with a generalized
selective measurement. Such maps for which $\sum_k E_kE_k^\dag\leq I$, describe processes in which extra information about what occurred in the process is
obtained by measurement. For non-trace-preserving quantum operations, $Tr[{\cal F}(\rho)]$ gives the probability
for the dynamical process ${\cal F}$ to occur, see Exercise 8.7 in \cite{Nils}.}.
Especially, in the cases like equation (\ref{AL}), where the quantum operation maps a density operator $\rho$, to another density $\varrho$, we have
$Tr[\varrho={\cal F}(\rho)]=Tr[\rho]=1$, which results in the completeness relation $\sum_k E_kE_k^\dag=I$. In this sense,
all the Kraus operators we consider in what follows, are trace-preserving quantum operation satisfying this closure relation.
One of the criteria that shows the effect of noise in the transmission
channel is the fidelity between the initial state of the channel and the (noisy) density matrix $\varrho$, that is,
\begin{equation}\label{AM}
F=\sqrt{\langle \Phi|\varrho|\Phi\rangle},
\end{equation}where $|\Phi\rangle$ in our model is the four-qubit cluster channel represented by the state (\ref{AC}). This quantity shows
how much information of the initial state will be affected by environmental effects \cite{Liang}. Clearly, if $F=1$, we have a perfect channel which
does not lose any information, while for $F=0$, all information will be destroyed. In our model, the initial density matrix of the channel is
\begin{equation}\label{AN}
\rho=|\Phi\rangle \langle\Phi|=\frac{1}{4}\left(|0000\rangle+|0011\rangle+|1100\rangle+|1111\rangle\right)
\left(\langle0000|+\langle0011|+\langle1100|+\langle1111|\right),
\end{equation}and hence the Kraus operators are of the form $E_{ijkl}=E_i\otimes E_j\otimes E_k\otimes E_l$. Under this
condition equation (\ref{AL}) takes the form
\begin{equation}\label{AN1}
\varrho=\sum_{i,j,k,l}E_{ijkl}\rho E^{\dag}_{ijkl},
\end{equation}in which the indices $i, j, k, l$ run over the number of the Kraus operators.
Here, we discuss the effect of some types of noise on the above channel:

\subsection{Bit-flip, phase-flip and bit-phase-flip channels}
These are some of the most common noises that have been investigated in the literature \cite{Nils,Mc}, see also \cite{Sin}. In a bit-flip channel
there is a probability $p$ for a bit-flip error, which means that state $|0\rangle$ becomes $|1\rangle$ and vice versa, while with the
probability $1-p$ nothing happens. On the other hand in the presence of a phase- flip channel the phases of the qubits will
be changed as $|0\rangle \rightarrow |0\rangle$ and $|1\rangle \rightarrow -|1\rangle$. There is also
another noise, bit-phase-channel, in which the bit and phase of the qubits change simultaneously. The operations
of these noises on the four-qubit channel
(\ref{AN}) can be described by the following Kraus operators
\begin{equation}\label{AN2}
E_0=\sqrt{1-p}I,\hspace{5mm}E_1=\sqrt{p}X_i,
\end{equation}where $i=1,2,3$ and the Pauli operators $X_1=X$, $X_2=Y$ and $X_3=Z$, stand for the bit-flip, phase-flip and bit-phase-flip noises respectively.
Using these operators in relation (\ref{AN1}), we get the following result
\begin{eqnarray}\label{AO}
\rho \mapsto \varrho=\left(\sqrt{1-p}I\right)^{\otimes 4} \rho \left(\sqrt{1-p}I\right)^{\otimes 4}+
\left(\sqrt{p}X_i\right)^{\otimes 4} \rho \left(\sqrt{p}X_i\right)^{\otimes 4}+\nonumber \\ \left[\left(\sqrt{1-p}I\right)^{\otimes 2}
\left(\sqrt{p}X_i\right)^{\otimes 2}\right]\rho \left[\left(\sqrt{1-p}I\right)^{\otimes 2}
\left(\sqrt{p}X_i\right)^{\otimes 2}\right]+\nonumber \\ \left[\left(\sqrt{p}X_i\right)^{\otimes 2}
\left(\sqrt{1-p}I\right)^{\otimes 2}\right]\rho \left[\left(\sqrt{p}X_i\right)^{\otimes 2}
\left(\sqrt{1-p}I\right)^{\otimes 2}\right]+...,
\end{eqnarray}where $...$ denotes twelve
other terms which since they do not contribute to the calculation of the fidelity, we omitted to write them explicitly. Therefore, we get
\begin{equation}\label{AP}
\varrho=\left[(1-p)^4+p^4+2p^2(1-p)^2\right]\rho+...,
\end{equation}which gives the following fidelity function
\begin{equation}\label{AR}
F=2p^2-2p+1.
\end{equation}
The behavior of this function shows that fidelity starts to decrease from
the value $F=1$ at $p=0$, until it reaches its non-zero minimum value in $p=1/2$, and then has an increasing behavior in the interval $1/2<p<1$.
Since the function is symmetric with respect to $p=1/2$, for each value of $p$
in the decreasing branch, there is a corresponding value in the increasing branch with the same value of fidelity.
So, in order
for the initial and final states to be as close as possible, it is necessary the value of $p$ to be as close to zero as possible.

\subsection{Depolarization channel}
In a depolarized noisy channel there is a probability $p$ that qubits are polarized, i.e., the principal system evolves into a
completely mixed state. This kind of noise leaves the system alone with probability $1-p$. The Kraus operators corresponding to such kind of noisy environment
are $E_0=\sqrt{1-p}I$ and $E_i=\sqrt{p/3}X_i$ with $1=1,2,3$, and $X_1=X$, $X_2=Y$ and $X_3=Z$, are Pauli matrices. Therefore, in depolarizing noisy
environment, the channel (\ref{AN}) evolves as

\begin{equation}\label{AS}
\rho\mapsto \varrho=\left(\sqrt{1-p}I\right)^{\otimes 4}\rho \left(\sqrt{1-p}I\right)^{\otimes 4}+\sum_{i=1}^3 \left(\sqrt{\frac{p}{3}}X_i\right)^{\otimes 4}
\rho \left(\sqrt{\frac{p}{3}}X_i\right)^{\otimes 4}+...,
\end{equation}where here, ... represents 252 other (cross) terms of the combination of $E_{ijkl}$ which
we did not express them explicitly in the above relation. This yields

\begin{equation}\label{AT}
\varrho=\left[(1-p)^4+\frac{p^4}{27}+...\right]\rho,
\end{equation}with fidelity function

\begin{equation}\label{AU}
F=\sqrt{(1-p)^4+\frac{p^4}{27}+...},
\end{equation}in which only the effects caused by $E_{iiii}$ are considered,
and for this, the following analysis of the behavior of the fidelity function will be an approximate analysis.
This function also has a decreasing branch in the interval $0<p<3/4$,
during which its value decreases from $F=1$ to its non-zero minimum at $p=3/4$ and then enters its increasing branch.
However, the value of the fidelity in this branch cannot exceed $F(1)=1/\sqrt{27}$.
Therefore, for values of the error parameter for which $F>1/\sqrt{27}$, the channel is subject to less damage.
This happens for a part of the decreasing branch of the fidelity function where the error parameter is placed in the interval $0<p<0.57$.

\subsection{Amplitude damping noise}
The process of amplitude damping is the result of energy loss in physical systems because of some types of interactions with the environment.
It can be shown that for a one qubit system the Kraus operators describing such a phenomenon are

\begin{equation}\label{AV}
E_0=\left(%
\begin{array}{cc}
1 & 0 \\
0 & \sqrt{1-p} \\
\end{array}%
\right),\hspace{5mm}E_1=\left(%
\begin{array}{cc}
0 & \sqrt{p} \\
0 & 0 \\
\end{array}%
\right),
\end{equation}in which $p$ represents the probability for the process $|1\rangle \rightarrow |0\rangle$, and the probability
that the system just stays in the state $|1\rangle$ is $1-p$. Therefore, in the presence of amplitude damping the
affected density matrix after the noise has been introduced in the channel (\ref{AN}) is
\begin{equation}\label{AW}
\rho\mapsto \varrho=\left(E_0^{\otimes 4}+E_0^{\otimes 2}E_1^{\otimes 2}+E_1^{\otimes2}E_0^{\otimes 2}+E_1^{\otimes 4}\right)\rho
\left(E_0^{\otimes 4}+E_0^{\otimes 2}E_1^{\dag \otimes 2}+E_1^{\dag \otimes2}E_0^{\otimes 2}+E_1^{\dag \otimes 4}\right)+....
\end{equation}
After some algebra we are led to the following form of the noisy channel
\begin{eqnarray}\label{AX}
\varrho=\frac{1}{4}\left[|0000\rangle+(1-p)|0011\rangle+(1-p)|1100\rangle+(1-p)^2|1111\rangle\right] \nonumber \\
\left[\langle0000|+(1-p)\langle0011|+(1-p)\langle1100|+(1-p)^2\langle1111|\right]\nonumber \\+
\frac{1}{4}\left[p|0000\rangle+p(1-p)|1100\rangle\right]\left[p\langle0000|+p(1-p)\langle 1100|\right]\nonumber \\+
\frac{1}{4}\left[p|0000\rangle+p(1-p)|0011\rangle\right]\left[p\langle0000|+p(1-p)\langle0011|\right]\nonumber \\+
\frac{p^4}{4}|0000\rangle\langle0000|+...,
\end{eqnarray}
where again ... denotes the remaining twelve cross terms whose contribution to evaluate the fidelity is zero.
The fidelity function in this case takes the form
\begin{equation}\label{AY}
F=\frac{1}{2}\sqrt{p^4-3p^3+7p^2-8p+4}.
\end{equation}It can be seen that the fidelity has a decreasing behavior in the interval $0<p<0.87$, at the end of which
it reaches to its minimum value and then increases until $F=1/2$ at $p=1$. Therefore, the channel has its best quality in the range $F> 1/2$.
Therefore, the highest efficiency of the channel is achieved when the value of the noise parameter is in the range
$0<p< 0.74$.

\subsection{Phase damping noisy channel}
Finally, let us introduce a quantum process during which the loss of information about relative phases in a quantum state occurs.
When phase damping happens, the channel interact with the environment according to the following Kraus operators

\begin{equation}\label{AZ}
E_0=\sqrt{1-p}I,\hspace{5mm}E_1=\left(%
\begin{array}{cc}
\sqrt{p} & 0 \\
0 & 0 \\
\end{array}%
\right),\hspace{5mm}E_2=\left(%
\begin{array}{cc}
0 & 0 \\
0 & \sqrt{p} \\
\end{array}%
\right),
\end{equation}where $p$ is the decoherence rate of phase damping, which describes the possibility of occurring error in quantum state due to travel qubit.
Under the act of this kind of noise the density matrix takes the form

\begin{eqnarray}\label{BA}
\rho\mapsto \varrho=\left(\sqrt{1-p}I\right)^{\otimes 4}\rho \left(\sqrt{1-p}I\right)^{\otimes 4}+\left(%
\begin{array}{cc}
\sqrt{p} & 0 \\
0 & 0 \\
\end{array}%
\right)^{\otimes 4}\rho \left(%
\begin{array}{cc}
\sqrt{p} & 0 \\
0 & 0 \\
\end{array}%
\right)^{\otimes 4}+\nonumber \\ \left(%
\begin{array}{cc}
0 & 0 \\
0 & \sqrt{p} \\
\end{array}%
\right)^{\otimes 4}\rho \left(%
\begin{array}{cc}
0 & 0 \\
0 & \sqrt{p} \\
\end{array}%
\right)^{\otimes 4}+...,
\end{eqnarray}in which $79$ cross terms are shown by ... as in the previous subsections.
A simple calculation shows that

\begin{equation}\label{BB}
\varrho=(1-p)^4 \rho+\frac{p^4}{4}\left(|0000\rangle \langle 0000|+|1111\rangle \langle 1111|\right)+...,
\end{equation}for which the fidelity becomes

\begin{equation}\label{BC}
F=\sqrt{(1-p)^4+\frac{p^4}{8}+...}.
\end{equation}An analysis similar
to what we did about the depolarization channel shows that if the parameter $p$ is
in the range $0<p<0.41$, the channel has less exposure to error. Note however that this is an estimate that is calculated
based on only the diagonal terms present in the fidelity function, and in this sense, it should be considered as an approximate interval for $p$.

Above, we have seen that a noisy channel can produce undesirable
effects. Now the question is that the errors resulting from the presence of noise can be corrected?
The development of the theory of quantum error correction is an important part of the theory of quantum computing,
since without it, we will not have a useful quantum computer.
Reviewing the correction of the errors described above is beyond the goals of this article, so we will only briefly look at some special cases.
In general, it can be shown that if the bit-flip, phase-flip and bit-phase-flip errors can be corrected, then any other type of error can also be corrected.
Here, we take a look at a technique known as the three qubit bit-flip code and its application to our channel.
Suppose that the only possibility for a noise to occur is flip of a qubit, which means that
during the channel the operator $X$ may act on a qubit.
To protect quantum states against the effects of such bit-flip noise, we encode a single qubit state in three qubits as
$|0\rangle \rightarrow |0_L\rangle=|000\rangle$ and $|1\rangle \rightarrow |1_L\rangle=|111\rangle$, \cite{Nils}. Here, we have used the notation $|0_L\rangle$
and $|1_L\rangle$, to emphasize that these are logical qubit states and not physical qubits. Thus, the logical form of the channel (\ref{AC}) takes the form

\begin{equation}\label{BD}
|\Phi\rangle_L=\frac{1}{2}(\bf{|0000\rangle_L+|0011\rangle_L+|1100\rangle_L+|1111\rangle_L}),
\end{equation}where $|{\bf 0\rangle_L}=|000\rangle$ and $|{\bf 1\rangle_L}=|111\rangle$,
which means that each term in the channel's state is encoded with a twelve-qubit state. Now, let us define the following thirteen projection operators

\begin{eqnarray}\label{BE}
\left\{
\begin{array}{ll}
P_0=|\bf{0000\rangle\langle0000|+|1111\rangle\langle1111|},\\
P_1=|{\it 100}{\bf 000}\rangle\langle{\it100}{\bf 000}|+|{\it 011}{\bf 111}\rangle\langle{\it011}{\bf 111}|,\\

P_2=|{\it 010}{\bf 000}\rangle\langle{\it010}{\bf 000}|+|{\it 101}{\bf 111}\rangle\langle{\it101}{\bf 111}|,\\

P_3=|{\it 001}{\bf 000}\rangle\langle{\it001}{\bf 000}|+|{\it 110}{\bf 111}\rangle\langle{\it110}{\bf 111}|,\\\\
...\\\\

P_{10}=|{\bf000}{\it 100}\rangle\langle{\bf 000}{\it 100}|+|{\bf 111}{\it 011}\rangle\langle{\bf 111}{\it 011}|,\\

P_{11}=|{\bf000}{\it 010}\rangle\langle{\bf 000}{\it 010}|+|{\bf 111}{\it 101}\rangle\langle{\bf 111}{\it 101}|,\\

P_{12}=|{\bf000}{\it 001}\rangle\langle{\bf 000}{\it 001}|+|{\bf 111}{\it 110}\rangle\langle{\bf 111}{\it 110}|.

\end{array}
\right.
\end{eqnarray}
It is possible to determine whether an error has occurred in any of the qubits by measuring the above projection operators. Indeed,
if the presence of noise has caused a bit-flip on the $i$th qubit, the outputs of the measurements (which are called error syndromes) will be

\begin{equation}\label{BF}
\langle\phi|P_i|\phi\rangle=1, \hspace{5mm}\langle\phi|P_{j\neq i}|\phi\rangle=0,
\end{equation}
where $|\phi\rangle$ is the state of the noisy channel. Obviously, the case $\langle\phi|P_0|\phi\rangle=1$, indicates that there are no errors at all.
After the location of the error qubit is determined, the error can be easily corrected from the desired state by applying the operator $X$ as

\begin{equation}\label{BG}
(I\otimes...\otimes I\otimes X \otimes I \otimes ...\otimes I)|\phi\rangle=|\Phi\rangle_L,
\end{equation}in which $X$ is placed in the $i$th location, where the corrupted qubit is located. It must be noted that
the syndrome measurement does not cause any change to the state: it is $|\phi\rangle$, both
before and after syndrome measurement. So, wee see that the error-correction procedure works good when the bit-flip occurs on one qubits.
If the bit-flip on a qubit being occurred with probability $p$, the error correction will occur with probability

\begin{eqnarray}\label{BH}
p_{ec}=(1-p)^{12}+12p(1-p)^{11}=1-p_e,
\end{eqnarray}where

\begin{equation}\label{BI}
p_e=1-\left[(1-p)^{12}+12p(1-p)^{11}\right],
\end{equation}is the probability of an error remaining uncorrected. This means that this protocol makes the transmission more reliable provided
$p_e< p$, which occurs whenever $p<0.017$, for which the fidelity (\ref{AR}) will be in the range $0.97<F<1$. We conclude this section
by noting that the method presented above can be extended to correct other types of errors such as phase-flip and
bit-phase-flip \cite{Nils,Mc}, although we will not go into more detail about them here.

\section{Summary}
In this paper, we have studied a BQT protocol in which Alice and Bob, the legitimate users, each own a $n$-qubit GHZ-like state
and send it to each other simultaneously. The channel they use to perform this protocol is a four-qubit cluster channel,
Alice and Bob each have two qubits of which.
To transmit $n$ qubits through this channel, at first, users convert their qubits into a simple single-qubit state, that
is a superposition of the states $|0\rangle$ and $|1\rangle$, and $n-1$ extra $|0\rangle$ qubits by applying a series of $\mbox{CNOT}$ gates.
Then, they perform their measurements on the state of the total system in the Bell basis and inform the
other the measurement results through a classical channel. We have shown that after these measurements,
the remaining states in the channel are collapsed into sixteen modes that provide the process of information transmission between users.
After Alice and Bob are informed of each other's measurement results, they reconstruct the initial single qubit state
that the other party intended to send, by applying appropriate unitary operators to the states in their hands. The final step of the protocol
is the conversion of these single-qubit to initial $n$-qubit states. It is shown that this cane be done by successively applying of a
series of $\mbox{CNOT}$ gates on the received states and some auxiliary qubits $|0\rangle$.

In the last part of the article, we have investigated some types of interactions of
the transmission channel with its environment, which causes noise in the channel and disrupts the teleportation process. The noises introduced are
bit-flip, phase-flip, bit-phase-flip, depolarization channel, amplitude damping noise and phase damping noisy channel.
For each of these cases, we have calculated the density matrix of the noisy channel and its fidelity function with the original channel.
Also, we extended the three qubit bit-flip code error correction technique to our model and presented an algorithm
based on which the channel can be protected from bit-flip noise.

\section*{Data Availability}
No data were used to support the authors' study in this
manuscript.

\end{document}